# Cryogenic Reconfigurable Logic with Superconducting Heater Cryotron: Enhancing Area Efficiency and Enabling Camouflaged Processors

Shamiul Alam[1], Dana S. Rampini[2], Bakhrom G. Oripov[2], Adam N. McCaughan[2], and Ahmedullah Aziz[1*]

[1]Department of Electrical Eng. & Computer Sci., University of Tennessee, Knoxville, TN 37996, USA
[2]National Institute of Standards and Technology, Boulder, CO 80305, USA
*Corresponding Author's Email: aziz@utk.edu

*Abstract*— Superconducting electronics are among the most promising alternatives to conventional CMOS technology thanks to the ultra-fast speed and ultra-high energy efficiency of the superconducting devices. Having a cryogenic control processor is also a crucial requirement for scaling the existing quantum computers up to thousands of qubits. Despite showing outstanding speed and energy efficiency, Josephson junction-based circuits suffer from several challenges such as flux trapping leading to limited scalability, difficulty in driving high impedances, and so on. Three-terminal cryotron devices have been proposed to solve these issues which can drive high impedances ($> 100\ k\Omega$) and are free from any flux trapping issue. In this work, we develop a reconfigurable logic circuit using a heater cryotron (hTron). In conventional approaches, the number of devices to perform a logic operation typically increases with the number of inputs. However, here, we demonstrate a single hTron device-based logic circuit that can be reconfigured to perform 1–input copy and NOT, 2–input AND and OR, and 3–input majority logic operations by choosing suitable biasing conditions. Consequently, we can perform any processing task with a much smaller number of devices. Also, since we can perform different logic operations with the same circuit (same layout), we can develop a camouflaged system where all the logic gates will have the same layout. Therefore, this proposed circuit will ensure enhanced hardware security against reverse engineering attacks.

Superconducting processors have garnered renewed interests in the past few decades primarily due to their promise as controllers for qubit programming/readout in large-scale quantum computing systems[1–5]. In addition, they are uniquely suited for exa-scale high-performance computing systems and space applications[6,7]. Josephson junction (JJ)-based logic families are the primary building blocks of superconducting Boolean logic. Thanks to the ultra-fast (>100 GHz) and ultra-low power (sub-aJ/bit switching energy) operation of JJ-based circuits[8,9], a superconducting processor has the potential to solve many existing issues of its CMOS counterpart. However, JJ-based circuits suffer from a number of challenges including difficulty in cascading, fabrication challenges, limited scalability due to flux trapping and sensitivity to magnetic field, and so on[10,11]. To solve some of these challenges, three-terminal cryotron devices were developed which show input gate current-driven switching of the channel between its superconducting and resistive states[10]. The heater-cryotron (hTron)[12] is a member of this cryotron family. These devices can drive large impedances ($> 100\ k\Omega$) and support a large number of fanouts due to their transition to a highly resistive state. Moreover, these devices do not require superconducting loops like JJs and hence, are free from flux trapping and resulting scalability issues[10]. hTron devices have already been used as an access device in cryogenic memories[13–15], as an interface between superconductors and semiconductors[12], to design logic circuits[10,11,16,17], in cryogenic neuromorphic systems[18–21], and so on.

Despite being used in several sensitive and critical applications, the hardware security techniques for superconducting processors are yet to be developed. Therefore, the superconducting processors are alarmingly vulnerable to different adversarial attacks including integrated circuit (IC) counterfeiting, IC masking, IC overproduction, intellectual property (IP) piracy, reverse engineering, etc. To prevent these attacks, different hardware security measures including IC camouflaging, logic locking, and/or adding a watermark, among others are used. Among these, logic locking[22] and IC camouflaging[23] have recently been utilized for developing secured superconducting hardware. In the logic locking technique, additional gates (AND, OR, XOR, etc.), key inputs, and additional on-chip memory are introduced into the original design to prevent external attacks. However, as reported in Ref.[22], the logic locking technique requires an additional 20% area overhead for only one OR gate. On the other hand, in IC camouflaging technique in Ref.[23], dummy JJs are introduced to make the layouts of all the gates look identical to prevent reverse engineering and other attacks. Both of these techniques have a significant negative impact on the area, delay, and power consumption of the circuit.

In this work, we demonstrate a superconducting reconfigurable logic circuit with a single hTron device that can perform the basic Boolean logic operations including 1–input copy and NOT, 2–input AND and OR, and 3–input majority. Typically, the number of devices required to perform any logic operation increases with the number of inputs[10,11]. However, in our proposed circuit, we perform all the above-mentioned single- and multi-input logic operations with only one hTron device. We apply different biasing conditions in the same circuit to perform all these logic operations. Thanks to the reconfigurability of the proposed circuit, it requires significantly less area compared to the existing superconducting logic circuits based on JJs, SQUIDs, and cryotrons. Moreover, as we are using the same circuit with the same layout to perform different tasks, we can build a camouflaged system (without compromising on any performance metric) where all the logic gates will look identical which can improve the hardware security of the system.

We start our discussion with the device characteristics of an hTron. hTron is a four-terminal current-driven superconducting device where two of the terminals form the gate and the other two form the superconducting channel (Figs. 1a–c). The gate and channel microwires are separated by a 25 nm SiO$_2$ dielectric spacer. The dielectric spacer electrically isolates but thermally couples the two microwires. Details on the device fabrication are available in the Supplement.

Initially, the channel remains superconducting for a given channel bias current ($I_{Ch}$) and no gate current ($I_G$). However, when $I_G$ is applied, the gate becomes resistive and generates thermal phonons which are carried to the channel by the dielectric spacer. The channel remains superconducting until $I_G$ exceeds a specific threshold and becomes resistive (Fig. 1b). However, when the gate becomes resistive, the presence of these thermal phonons suppress superconductivity in the channel, and the more gate current is added, the greater the suppression of $I_{Ch}^C$ is observed (as shown in Fig. 1d). The reason of suppressing the superconductivity is that the thermally-generated phonons are capable of breaking

Cooper pairs in the superconducting channel (Fig. 1c). When $I_G$ exceeds a specific threshold ($I_G^C$) such that the channel critical current ($I_{Ch}^C$) (a function of $I_G$, shown in Fig. 1d) becomes smaller than the applied $I_{Ch}$, enough phonons with sufficient energy ($> 2\Delta$, where $\Delta$ is the superconducting energy gap[24]) are generated which can break Cooper pairs in the channel and cause the entire channel to switch to the high impedance resistive state, driving $I_{Ch}$ to the external circuitry (Fig. 1c).

While characterizing the device, we first apply a fixed current to the gate ($I_G$). Then we ramp up the channel bias current ($I_{Ch}$) from 0 A to above $I_{Ch}^C$ so that the channel switches. For each $I_G$, we perform the ramping of $I_{Ch}$ 50 times and record the $I_{Ch}^C$ for each measurement. Figure 1d shows the measured distribution of $I_{Ch}^C$ for $I_G$ ranging from 10 $\mu A$ to 135 $\mu A$. We also extract the median of $I_{Ch}^C$ for each $I_G$ (Fig. 1e) and use that as the nominal $I_{Ch}^C$ in our circuit design. Next, we perform a transient measurement to show the time dynamics of the channel switching from its superconducting to non-superconducting state. The inset of Fig. 1f shows the transient measurement setup with a load resistance of 1 $k\Omega$. Here, we first ramp $I_{Ch}$ up to 55 $\mu A$ and then start ramping up $I_G$ to capture the gate-driven switching (Fig. 1f). As shown in Fig. 1g, initially the channel is in its superconducting state and hence, no current flows through the load resistor ($I_L = V_L \approx 0$). However, when $I_G$ exceeds $I_G^C$ (110 $\mu A$ for this case), the channel switches to its non-superconducting state and drives $I_{Ch}$ to the load resistor which creates a nonzero voltage across it. The turn-on and turn-off times for the hTron device considered in this work are 300 ps and 15 ns, respectively. However, the turn-on and turn-off time can be reduced by using higher input energy and different superconducting material, respectively[12]. The WSi film used as the superconducting material had a critical temperature ($T_C$) of 3.5 K and all the measurements were performed at 0.9 K. To obtain a higher operating temperature, a different material such as NbN or NbTiN can be used which show a $T_C$ of 9-12 K depending on the stoichiometry. More details on the device characterization are available in the Supplement.

In this work, we utilize the gate current-controlled superconducting to non-superconducting transition of the hTron to design a reconfigurable logic circuit for achieving secured superconducting hardware. We use the same measurement setup (shown in the inset of Fig. 1f) and measured device characteristics (Fig. 1e) to design the logic circuit. The only difference is that instead of one input current to the gate, we use three input currents to perform up to three-input logic operations. Figure 2a shows the schematic of the designed reconfigurable logic circuit where $I_{IN1}$, $I_{IN2}$, and $I_{IN3}$ are the input currents corresponding to the number of inputs of the logic operations, $I_{B1}$ is a common gate bias used for all the operations, and $I_{B2}$ is the channel bias. In our design, we define the two logic states ('0' and '1') with two current levels (0 $\mu A$ and $I_G^C/2$, respectively) (Fig. 2b). Figure 2c shows the biasing scheme that we use to perform different logic functions with the same circuit shown in Fig. 2a. For all the logic operations, $I_{B2}$ is kept fixed at 55 $\mu A$ (chosen based on the measured device data shown in Figs. 1d–g). Now, to perform different logic functions such as 1–input copy and NOT, 2–input AND and OR, and 3–input majority, we only need to apply different bias currents to $I_{B1}$ (range is shown in Fig. 2c). The combination of the input currents and $I_{B1}$ will determine the switching of the channel under the fixed $I_{B2}$ bias and accordingly, we will get zero or nonzero current at the output.

To demonstrate the functionality of different logic functions with our proposed circuit, we adopt a simulation–based approach. We first develop a lookup–table (LUT) based compact model for the hTron in Verilog–A and calibrate the model with the measured data shown in Figs. 1d–g. In each time step, the model takes applied gate and channel currents as input and based on the value of the gate current, reads the corresponding $I_{Ch}^C$ from the LUT. Finally, the model compares the applied channel current with the $I_{Ch}^C$ obtained from the LUT and determines whether the channel will switch or not. As already mentioned, our proposed circuit is similar to the measurement setup. The only difference is the use of multiple input currents applied to the gate. Note, all the input currents and $I_{B1}$ are added together and seen as one bias current by the gate. Therefore, utilizing the experimentally calibrated device model, our simulation replicates the behavior that we would observe in the experiment.

First, we discuss the 1–input copy gate. Here, we use the $IN1$ terminal to apply the input current and we apply 65 $\mu A$ as $I_{B1}$ and 55 $\mu A$ as $I_{B2}$. As shown in Fig. 3a, for a logic '0' at the input ($I_{IN1} = 0$), the total applied current to the gate is $I_{B1}$ (65 $\mu A$) which is less than $I_G^C$ (110 $\mu A$) and hence, the channel remains superconducting. Therefore, $I_{B2}$ flows through the channel and we get logic '0' ($I_{OUT} = 0$) at the output. On the other hand, as shown in Fig. 3b, when a logic '1' is applied ($I_{IN1} = I_G^C/2 = 55\ \mu A$), the total gate current ($I_G = 120\ \mu A$) exceeds $I_G^C$ and switches the channel to its non-superconducting state which is highly resistive. Therefore, $I_{B2}$ flows through the external resistor ($R = 1\ k\Omega$) and we get logic '1' ($I_{OUT} = I_G^C/2 = 55\mu A$) at the output. Figures 3c-e show the simulated time dynamics of the copy gate which verifies the functionality of the logic function.

Next, we discuss the 1–input NOT gate where we use the same input terminal and same $I_{B2}$. The only difference is that we apply $-120\ \mu A$ as $I_{B2}$. Now, when a logic '0' ($I_{IN1} = 0$) is applied, only $I_{B1}$ flows through the gate which itself is greater than $I_G^C$ (Fig. 3f). As a result, the channel of the device switches to the non-superconducting state and drives $I_{B2}$ to the external resistor. Therefore, we get logic '1' ($I_{OUT} = I_G^C/2 = 55\mu A$) at the output. However, when the input is logic '1' ($I_{IN1} = 55\ \mu A$), the device gets a total gate current of $-65\ \mu A$ which cannot switch the channel (Fig. 3g). Hence, the channel remains superconducting and we get 0 A at the output. Figures 3h–j verify the NOT gate functionality.

With the same circuit, we can also perform 2–input Boolean logic functions such as AND and OR logic where we use two input terminals to apply the input currents ($I_{IN1}$ and $I_{IN2}$).

To perform AND operation, we apply $I_{B1} = 10\ \mu A$. As a result, for input combinations of '00' and '01' (or '10'), the gate gets 10 $\mu A$ and 65 $\mu A$, respectively which are not sufficient enough to switch the state of the hTron channel (Fig. 4a). Therefore, the channel remains in its superconducting state and drives 0 $\mu A$ (logic '0') to the output. On the other hand, for the '11' input combination, a total current of 120 $\mu A$ flows through the gate and switches the channel to the non-superconducting state (Fig. 4b). Therefore, $I_{B2}$ flows through the external resistor and we get logic '1' (55 $\mu A$) at the output. The time dynamics of AND functionality are shown in Figs. 4c–e.

Next, to perform the OR operation, we apply $I_{B1} = 65\ \mu A$. As a result, whenever one of the two inputs is logic '1', the gate terminal gets sufficient current to switch the channel of the device. Therefore, only for the input combination of '00', the channel remains superconducting and for '01', '10', and '11' combinations, the channel switches to its non-superconducting state (Figs. 4f,g). As seen in Figs. 4h–j, we get the OR operation.

Majority is a Boolean logic operation that generates an output of logic '1' if there are more number of logic '1' among an odd number of inputs. The majority logic can be utilized to perform the basic logic operations such as AND, OR, XOR, etc., and requires a lower number of gates to develop arithmetic-intensive systems[25]. Compared to the AND implementation, majority reduces the logical depth of the arithmetic-intensive circuits by up to 33%[26]. For example, while building a 1–bit full adder, the use of majority can reduce the logic level by 50% and 57.14% compared to the NAND and NOR implementations, respectively, and can reduce the number of required cycles by 60%[26].

Here, we show that the same circuit with a single hTron device can be reconfigured to perform 3–input majority logic operation. Here, we use $I_{B1} = 10\ \mu A$. As a result, when there is less number of logic '1' among the inputs, the gate terminal does not get sufficient current to switch the channel (Fig. 5a). On the other hand, when there is more number of logic '1' inputs, the total gate current exceeds $I_G^C$ and the channel switches to its non-superconducting state (Fig. 5b). As seen in the transient results of Figs. 5c-e, the same circuit can perform the 3–input majority logic function.

We demonstrate a single hTron-based reconfigurable logic circuit that can perform 1–input copy and NOT, 2–input AND and OR, and 3–input majority operations by only choosing suitable biasing conditions. Our proposed logic circuit benefits from useful features of the hTron including low turn-on-time, input-output isolation, low power operation, and so on[12]. Moreover, hTron devices do not need any superconducting loop like other superconducting devices such as JJs and SQUIDs. Therefore, our proposed logic circuit will not suffer from any flux trapping issue and will be able to solve the scalability issue of the JJ and SQUID-based circuits. Typically, the number of required devices in a logic gate increases with the number of inputs which is not the case for our design. We use the same circuit based on one hTron device to reconfigurably perform all these basic operations, enabling the implementation of a superconducting processing unit with a much less number of devices compared with the existing approaches. Moreover, since we are using the same circuit for different logic operations, we can develop a camouflaged processing unit where all the logic gates will have the same layout and will reduce the risk of reverse engineering.

Also, thanks to the high impedance ($\gg 1\ k\Omega$) non-superconducting state of hTron devices, the cascadability of the proposed logic circuit is not a concern like the other current-controlled superconducting logic gates. One hTron can drive other resistive gate terminals (typically has an impedance of $1\ k\Omega$) of the next stage logic circuits. Additionally, our proposed copy gate can be used to develop a current splitter circuit like the one proposed in Ref.[27] to enable larger fanout.

Finally, hTron devices suffer from reset time limitations due to the thermal recovery after each switching and so will be the case for our proposed circuit. To solve this issue, a different superconductor with a suitable thickness (for example, NbN with $\sim 1\ ns$ thermal reset time[28]) or the same superconductor with different thickness can be used to make the channel microwire. Furthermore, our proposed circuit is not limited to thermal hTron devices. Our proposal will be applicable to any device that shows gate-controlled switching between superconducting and non-superconducting states like Josephson junction FET[16,29], dayem transistor[30,31], ferroelectric SQUID[11,15], and so on.

**Supplementary Material**

See Supplemetary Material for the detailed discussion on fabrication and electrical characterization of the heater cryotron (hTron) device.

**Acknowledgment**

This research was funded by the University of Tennessee Knoxville (https://ror.org/020f3ap87) and NIST (https://ror.org/05xpvk416).The U.S. Government is authorized to reproduce and distribute reprints for governmental purposes notwithstanding any copyright annotation thereon.

**Data Availability**

The data that support the plots within this paper and other findings of this study are available from the corresponding author upon reasonable request.

**Author Contributions**

S.A. conceived the idea, designed the logic gates, performed the simulations, and prepared the draft manuscript. B.G.O. fabricated the hTron device. D.S.R. performed the measurements for the fabricated hTron device. A.N.M. and A.A. analyzed, helped finalize the designs, and edited the draft manuscript. A.A. supervised the project.

**Competing Interests**

The authors declare no competing interests.

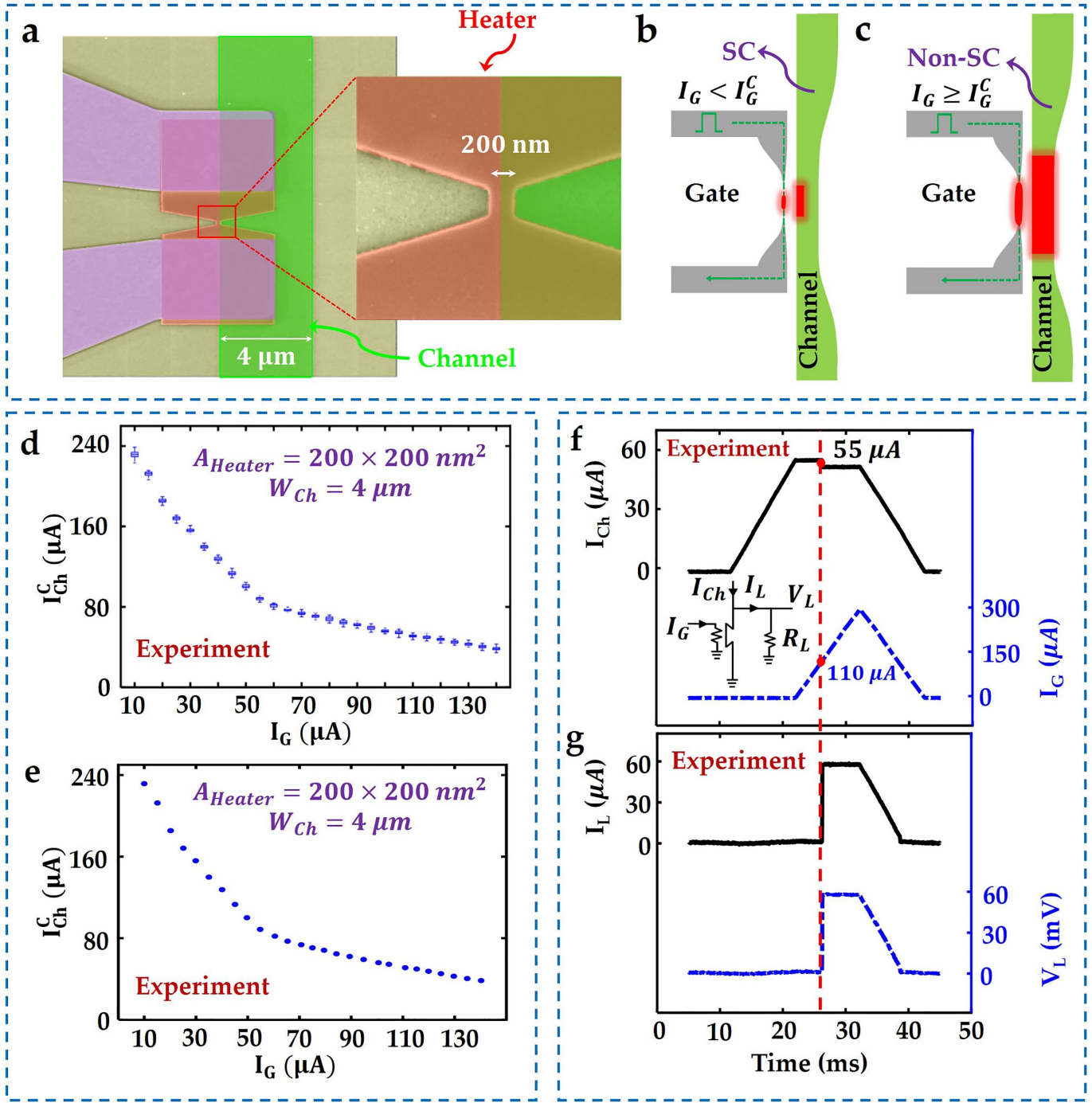

**Fig. 1. Device characteristics of heater cryotron (hTron). (a)** False-colored scanning electron micrograph (SEM) of a fabricated hTron device. The device consists of a resistive heater (gate) and a superconducting microwire (channel) separated by a SiO₂ dielectric spacer. The fabricated device has a heater gate with an area of $200 \times 200\ nm^2$ and a channel with a width of $4\ \mu m$. **(b), (c)** Illustration of the switching mechanism of a hTron device. **(d)** Measured distributions of the channel critical current ($I_{Ch}^C$) for different values of $I_G$ (please refer to Fig. S1 in the supplement for the zoomed-in views of the measured $I_{Ch}^C$ distribution). **(e)** Extracted median values of $I_{Ch}^C$ distributions for each $I_G$. **(f),(g)** Measured time dynamics of the switching of a hTron device with a load resistance ($R_L$) of $1\ k\Omega$.

Cryogenic Reconfigurable Logic with Superconducting Heater Cryotron: Enhancing Area Efficiency and Enabling Camouflaged Processors

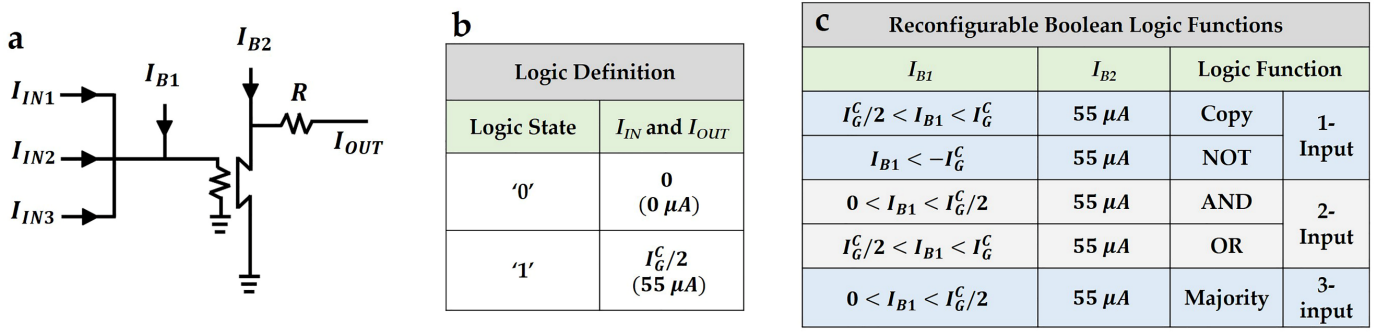

**Fig. 2. Single hTron-based reconfigurable logic circuit. (a)** Schematic of the hTron-based reconfigurable logic circuit that can perform 1-input copy and NOT, 2-input AND and OR, and 3-input majority operations with only a suitable bias current ($I_{B1}$). We use 1 $k\Omega$ as the load resistor ($R$). **(b)** Definition of logic states. **(c)** Biasing conditions to perform different logic operations with the same circuit.

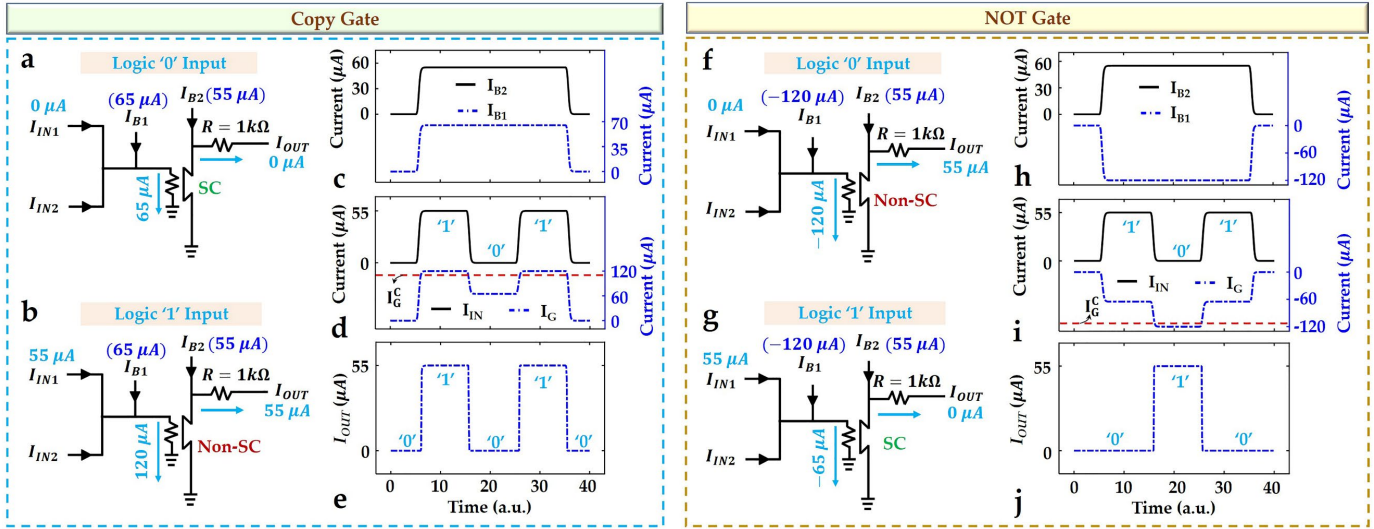

**Fig. 3. 1-input copy and NOT operations.** Copy operation with the proposed circuit when the input is **(a)** logic '0' and **(b)** logic '1'. **(c)-(e)** Simulated time dynamics of the copy operation. **(f),(g)** Illustration of the NOT operation for logic '0' and '1' inputs, respectively. **(h)-(j),** Time dynamics of the NOT operation where the channel switches for logic '0' input due to the choice of $I_{B1}$.

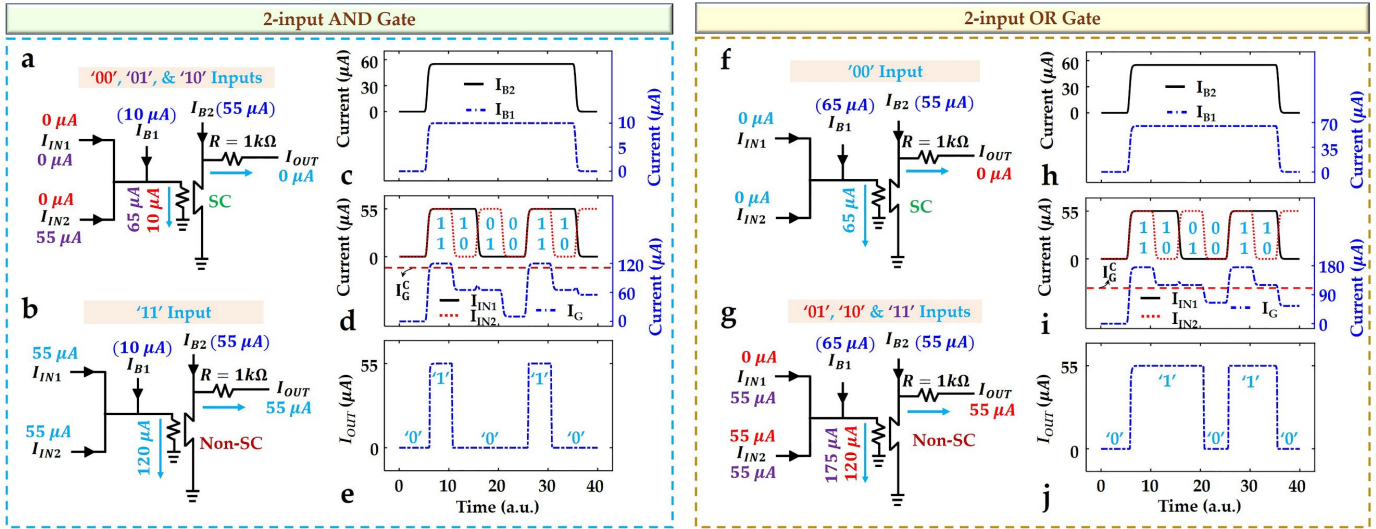

**Fig. 4. 2-input AND and OR operations.** AND operation with the proposed reconfigurable circuit for (**a**) '00', '01', and '10' and (**b**) '11' input combinations. (**c**)-(**e**) Simulated time dynamics for the AND operation. (**f**),(**g**) Illustration of the OR operation for '00' and '01', '10' and '11' input combinations, respectively. (**h**)-(**j**) Time dynamics of the OR operation where the channel switches for the '11' combination due to the choice of $I_{B1}$.

Cryogenic Reconfigurable Logic with Superconducting Heater Cryotron: Enhancing Area Efficiency and Enabling Camouflaged Processors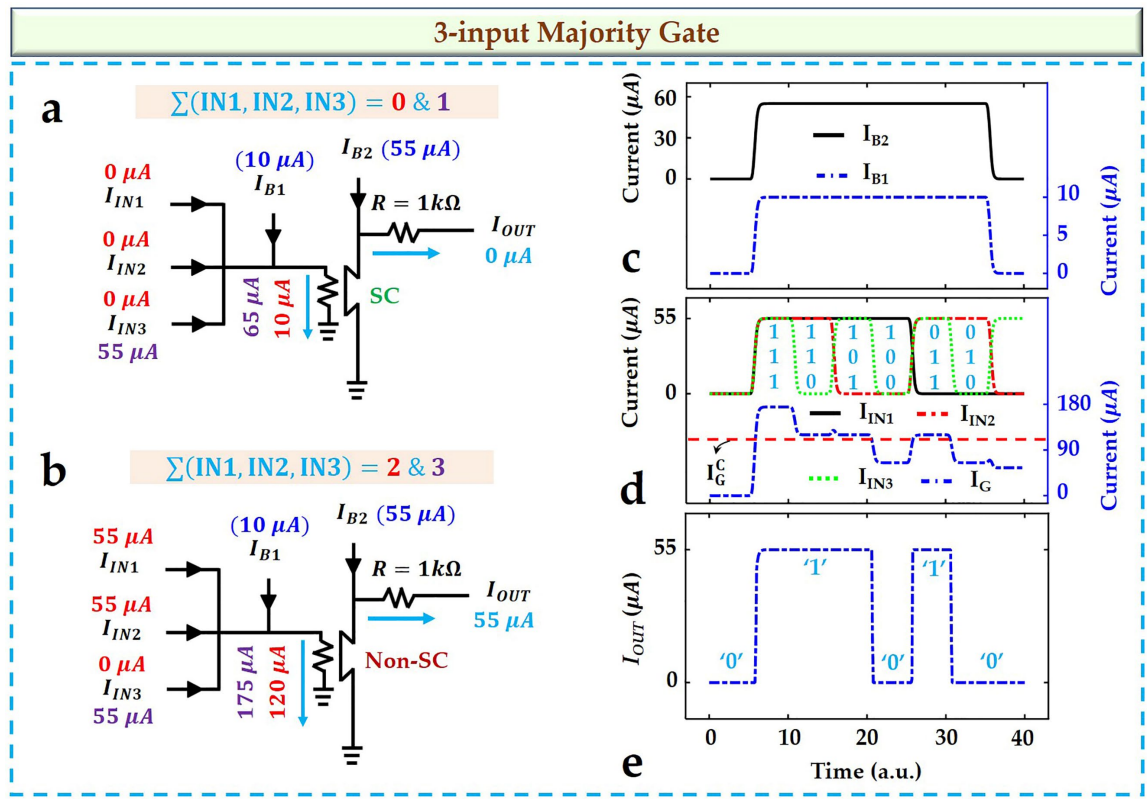

**Fig. 5. 3-input Majority operation.** Illustation of majority logic when there are (**a**) more and (**b**) less number of logic '1' inputs. (**c**)-(**e**) Simulated time dynamics of the Majority operation.

# Supplementary Material for
# Cryogenic Reconfigurable Logic with Superconducting Heater Cryotron: Enhancing Area Efficiency and Enabling Camouflaged Processors

Shamiul Alam[1], Dana S. Rampini[2], Bakhrom G. Oripov[2], Adam N. McCaughan[2], and Ahmedullah Aziz[1*]

[1]Department of Electrical Eng. & Computer Sci., University of Tennessee, Knoxville, TN 37996, USA
[2]National Institute of Standards and Technology, Boulder, CO 80305, USA
*Corresponding Author's Email: aziz@utk.edu

1. **Device fabrication**

The devices described here were fabricated using a combination of electron beam lithography and optical lithography. First, a layer of superconducting tungsten silicide (WSi) with a thickness of 4 $nm$ was deposited via magnetron sputtering on a silicon wafer. Contact pads of 5 $nm$ Ti followed by 30 $nm$ Au were deposited using a liftoff process and optical lithography. The WSi was then patterned using electron beam lithography and etched in an RIE with an $SF_6$ chemistry to form the channel of the device. For the dielectric spacer layer that separates the heater from the channel, 25 $nm$ of $SiO_2$ was deposited via sputtering over the entire wafer. After this, another WSi layer with the same thickness of 4 $nm$ was deposited via sputtering and once again patterned and etched to form the heater input gate for the device. Contact pads were added to this upper layer with added with the same liftoff process as the first layer. Lastly, openings were etched through the $SiO_2$ dielectric using an RIE with a $CHF_3:O_2$ chemistry so that electrical contact could be made with the bottom (channel) layer.

2. **Device characterization**

The measurements made on the devices here were taken using an arbitrary waveform generator (AWG) with two channels and 10 $k\Omega$ resistors in series with each channel. For each datapoint, one channel of the AWG was configured to a set voltage so that it delivered a fixed gate current, and the other channel was configured to ramp up from zero current to above the device channel switching current. The voltage of the device channel was recorded during this process on an oscilloscope, and when it exceeded a threshold of 50 $mV$ the device was determined to have transitioned out of the superconducting state, and that channel bias current was recorded as the channel switching current for that measurement. This process was repeated 50 times for each gate current (as shown in Fig. S1), and the median value of those 50 measurements was extracted as the device channel switching current for that gate current.

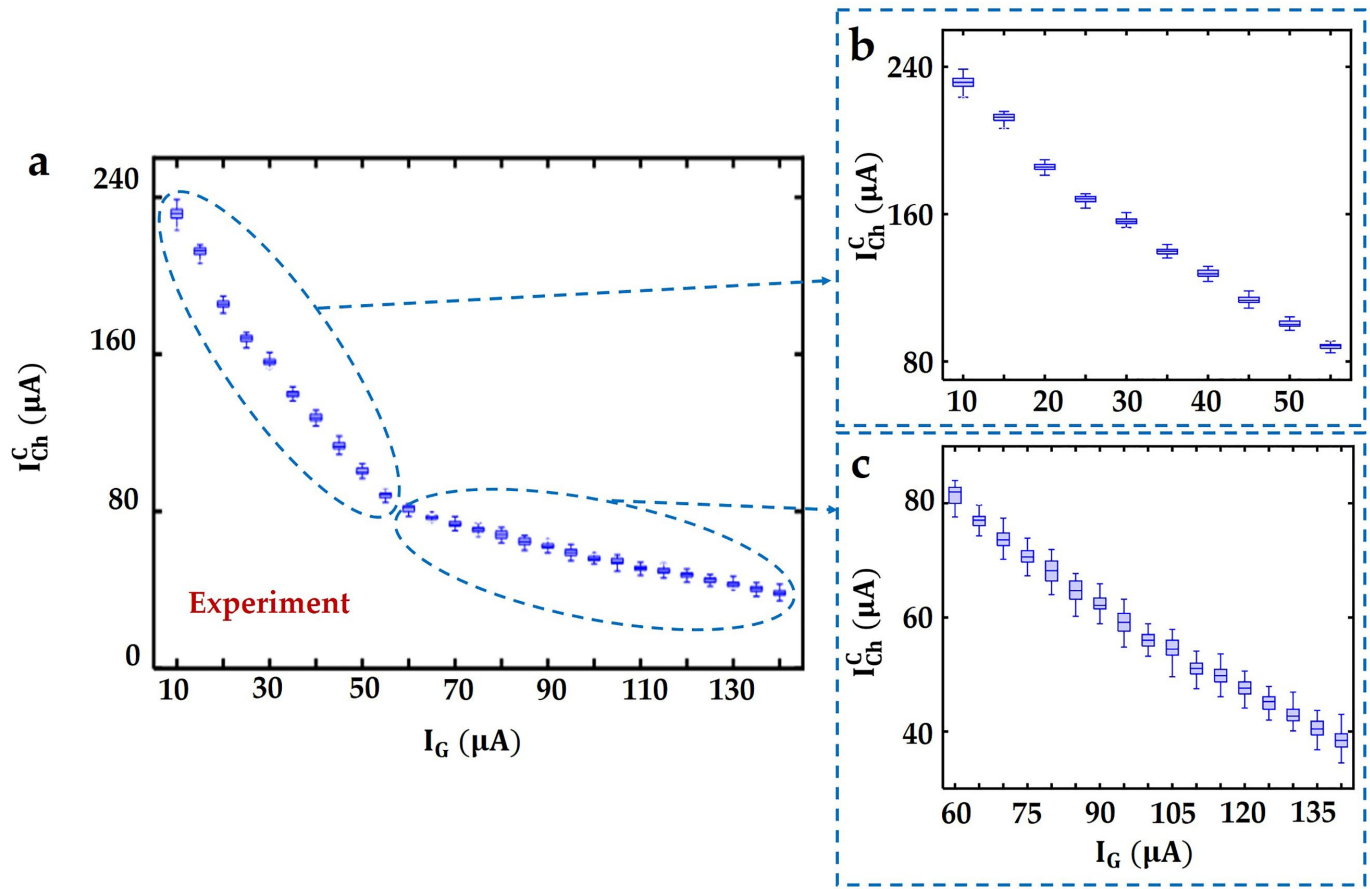

**Fig. S1. (a)** Measured distributions of the channel critical current ($I_{Ch}^{C}$) for different values of $I_G$. **(b) – (c)** Zoomed-in views of the measured $I_{Ch}^{C}$ distribution shown in (a).